\documentclass[10pt,aps,prd,nofootinbib,twocolumn,superscriptaddress,noeprint,nolongbibliography]{revtex4-2}

\emergencystretch=\maxdimen
\usepackage{graphicx}
\usepackage{dcolumn}
\usepackage{amsmath}
\usepackage{amssymb}  
\usepackage{mathtools}
\usepackage{float}
\usepackage{pgfplots}
\usepackage{booktabs}
\usepackage{bm}
\usepackage{multirow}
\usepackage{epsfig}   
\usepackage{tensor} 
\usepackage[colorlinks=true, citecolor=blue, urlcolor = blue, linkcolor= red, bookmarks=true]{hyperref}

\begin{document}
\title[]{Testing Strong Gravitational Lensing Effects of Supermassive Black Holes with String-Inspired Metric: Observational Signatures and EHT Constraints}

\author{Amnish Vachher}\email{amnishvachher22@gmail.com} 
\affiliation{Centre for Theoretical Physics, 
	Jamia Millia Islamia, New Delhi 110025, India}
\author{Shafqat Ul Islam} \email{shafphy@gmail.com}
\affiliation{Astrophysics and Cosmology Research Unit, 
	School of Mathematics, Statistics and Computer Science, 
	University of KwaZulu-Natal, Private Bag 54001, Durban 4000, South Africa}
\author{Rahul Kumar Walia}\email{rahulkumar@arizona.edu}
\affiliation{Department of Physics, University of Arizona
1118 E. Fourth Street, Tucson, AZ 85721, United States}
\author{Sushant~G.~Ghosh }\email{sghosh2@jmi.ac.in}
\affiliation{Centre for Theoretical Physics, 
	Jamia Millia Islamia, New Delhi 110025, India}
\affiliation{Astrophysics and Cosmology Research Unit, 
	School of Mathematics, Statistics and Computer Science, 
	University of KwaZulu-Natal, Private Bag 54001, Durban 4000, South Africa}
 
\begin{abstract}
We analyze gravitational lensing in the strong field limit for spherically symmetric string-inspired Euler-Heisenberg black holes, characterized by magnetic charge ($q$) and Einstein-Maxwell-dilaton coupling constants ($\alpha, \beta$) from the low-energy limit of heterotic string theory. Our results show that the string coupling has a weak impact on the positions of relativistic images, deflection angles, photon orbit radii, and shadow sizes, making these black holes indistinguishable from the Gibbons-Maeda-Garfinkle-Horowitz-Strominger (GMGHS) black holes with the same mass and charge. Compared to Schwarzschild black holes, the string-inspired Euler-Heisenberg black holes exhibit smaller deflection angles, decreasing with increasing charge. Moreover, the time delay for Sgr A * and M87 * can reach $~11.477$ and $~17349.8$ minutes, respectively, at $q=0.1$ and $\eta=-1$, deviating from Schwarzschild black holes by $~0.0198$ and $~28.9$ minutes, which are not very significant. For Sgr A* and M87*, we determine $\theta_\infty$ range within $(11.52, 26.33)~\mu as$, and $(9.17, 19.78)~\mu as$ respectively, with angular separations $s$ ranging from $(3.29-6.85)~nas$ for Sgr A* and $(2.47-5.15)~nas$ for M87*. EHT bounds on the $\theta_{sh}$ of Sgr A* and M87* within the $1\sigma$ interval bound the $q$ as: for Sgr A* $0.54109\le q \le 0.7796 $ and for M87* $0< q \le 0.29107$, while in both the cases, we did not find any bound on the parameter $\eta$. We show that string-inspired Euler-Heisenberg black holes and EHT observations agree in the finite parameter space. A discussion on the effective metric has been included.
\end{abstract}
\maketitle

\section{Introduction}\label{intro}
The black hole shadow observations \cite{EventHorizonTelescope:2019dse,EventHorizonTelescope:2019pgp,EventHorizonTelescope:2019ggy,EventHorizonTelescope:2022wkp,EventHorizonTelescope:2022urf,EventHorizonTelescope:2022xqj,EventHorizonTelescope:2022xqj,EHT:2023ujh,Collaboration:2024unf} by the Event Horizon Telescope (EHT) collaboration have opened up an exciting arena for understanding gravity in extreme and relativistic field regimes. These observations have significantly reinforced the validation of general relativity (GR) in the strongest-field regime and shed light on the captivating phenomenon of strong gravitational lensing of hot accretion flow around supermassive black holes. Gravitational lensing, as predicted by GR, is now an indispensable tool for understanding the structure of spacetime, studying astrophysical objects, especially black holes, and understanding the dynamics of galaxy clusters and the detection of dark matter \cite{Einstein:1936llh, Liebes:1964zz,Mellier:1998pk,Bartelmann:1999yn,Schmidt:2008hc,Barnacka:2013lfa}. A light passing close to a compact astrophysical object is gravitationally lensed, offering a unique opportunity to study gravity on a large scale and phenomena that would otherwise remain hidden from direct observation.

The exploration of gravitational lensing has a rich history, dating back to the inception of GR itself \cite{Perlick:2010zh}. In particular, Darwin \cite{unknown-author-no-date} was among the first to study the gravitational lensing concept, and later, Luminet~ \cite{Luminet:1979nyg} studied it in the context of Schwarzschild black holes. Virbhadra and Ellis \cite{Virbhadra:1999nm, Virbhadra:2002ju} studied the strong-field gravitational lensing phenomena and developed a framework to investigate the formation and position of relativistic images around Schwarzschild black holes. In their study, Virbhadra and Ellis \cite{Virbhadra:1999nm} not only identified primary and secondary images but also discovered a series of relativistic images produced by photons making several loops around the black hole on either side of the optic axis. Later, Frittelli {et al.}\cite{Frittelli:1999yf} provided more rigorous analytical descriptions of the lens equation and its solution, comparing their findings with those of Virbhadra and Ellis. Subsequently, Bozza \cite{Bozza:2001xd, Bozza:2002zj, Bozza:2007gt, Bozza:2002af} and Tsukamoto \cite{Tsukamoto:2016jzh, Tsukamoto:2017fxq} introduced novel methods to study strong lensing in a parameterized spherically symmetric static spacetime and showed that the deflection angle diverges logarithmically as light rays approach the photon sphere. Torres \cite{Eiroa:2002mk} derived analytical expressions for the positions and magnifications of relativistic images in Reissner-Nordstr\"{o}m (RN) black hole lensing. These methods have significantly contributed to our understanding of strong-field gravitational lensing and have gained general acceptance in the scientific community to test the predictions of GR \cite{Bozza:2010xqn, Eiroa:2002mk, Iyer:2006cn, Tsukamoto:2016jzh, Virbhadra:2007kw, Shaikh:2019jfr}. They are also frequently applied to black holes in modified gravity theories~\cite{Ghosh:2020spb, Whisker:2004gq, Abbas:2021whh, Eiroa:2005ag, Gyulchev:2006zg, Ghosh:2010uw, Gyulchev:2012ty, Molla:2023hog, Grespan:2023cpa, Kumar:2022fqo, Kumar:2021cyl, Lu:2021htd, Ali:2021psk, Hsieh:2021scb,Chen:2009eu,Sarkar:2006ry,Javed:2019qyg,Shaikh:2019itn,Eiroa:2010wm,Ovgun:2019wej,Panpanich:2019mll,Bronnikov:2018nub,Shaikh:2018oul, Lu:2021htd,Babar:2021nst}, including higher curvature gravity theories \cite{Kumar:2020sag,Islam:2020xmy,Narzilloev:2021jtg}, to estimate black hole parameters \cite{Afrin:2023uzo,Afrin:2021imp}, and to determine any matter distribution in a black hole background \cite{Vachher:2024ldc,Gao:2023ltr,2795690}. In particular, strong-gravitational lensing near black holes unveils phenomena such as relativistic images, Einstein rings, shadows, and photon rings~\cite{Synge:1966okc, Gralla:2019xty, darwin1959gravity, Cunha:2018acu, Bozza:2010xqn, Bozza:2001xd, Bozza:2002zj, Bozza:2002af}.

Gravitational lensing has also been employed to test the strong-field predictions of black holes from string theories \cite{Younesizadeh:2022czv,Molla:2024lpt,Sharif:2017ogw}, loop quantum gravity \cite{Dong:2024alq,Fu:2021fxn,Sahu:2015dea,KumarWalia:2022ddq,Kumar:2023jgh,Islam:2022wck}, and supergravity theories \cite{ALI2024}, to investigate whether there are any measurable differences from those of a Schwarzschild black hole. Bhadra \cite{Bhadra:2003zs} investigated strong gravitational lensing for charged black holes from heterotic string theory (GMGHS black holes)~\cite{Gibbons:1987ps, Garfinkle:1990qj} and found no significant string effects on observable parameters, making it difficult to distinguish them from RN black holes with the same charge. However, Kerr-Sen black holes, the rotating counterparts of GMGHS black holes, exhibit distinctive lensing characteristics, such as larger deflection angles, bigger and more deformed shadows, and shifts in relativistic caustics compared to Kerr-Newman black holes \cite{Xavier:2020egv,Gyulchev:2006zg}. In the low-energy limit of string theory, the dilaton parameter significantly increases the size of the black hole shadow and the position of the relativistic image \cite{Younesizadeh:2022czv}. Molla et al.~\cite{Molla:2024lpt} examined the gravitational lensing behavior of spherically symmetric $\alpha$-corrected RN black holes derived from heterotic superstring effective field theory. These results highlight the potential of gravitational lensing to probe the underlying structure of spacetime and validate the predictions of string theory, making it a powerful tool in the search for new physics.

Inspired by the aforementioned context, this study focuses on the strong-gravitational lensing produced by the string-inspired Euler-Heisenberg black holes \cite{Bakopoulos:2024hah}. It compares these predictions with those from the GMGHS black holes \cite{Bhadra:2003zs} and the Schwarzschild black holes \cite{Bozza:2002zj}. Recently,  Bakopoulos et al.~\cite{Bakopoulos:2024hah} considered a gravitational theory of a non-linear electrodynamics (NED) field non-minimally coupled with a dilaton field with a specific non-trivial coupling, and obtained an analytical black hole solution which is a generalization of the GMGHS black hole, where a linear electrodynamics field is replaced by a Euler-Heisenberg NED field. This model has been investigated to discuss shadows and quasi-normal modes \cite{Huang:2025jfa,Lambiase:2024lvo} and particle motion in modified surroundings \cite{Mustafa:2025lix,Su:2024lvs}. We analyze the positions, separations, magnifications, and time delays in the formation of relativistic images of these black holes and estimate their values for the EHT targets, supermassive black holes, Sgr A* and M87*. Specifically, we examine how the dilaton coupling parameters, compared to those affected by the GMGHS black hole, affect various gravitational lensing observables and the time delay observed between relativistic images. 

The paper is organized as follows: The spacetime structure of string-inspired Euler-Heisenberg black holes is briefly reviewed for completeness, with a particular focus on identifying parameter space of solutions for black holes and no-black holes in Section~\ref{solution}. The framework for gravitational lensing, including the lens equation, deflection angle, and coefficients for strong lensing, is outlined in Section~\ref{lensing}. We analyze strong-gravitational lensing observables in Section~\ref{observation}, which include the position of the innermost image ($\theta_{\rm \infty}$), the separation between images ($s$), the flux ratio of the first image to all others ($r_{\rm mag}$), and the time delay between the first and second relativistic images for Sgr A* and M87*. Section~\ref{constraint} focuses on constraining and estimating parameters for the string-inspired Euler-Heisenberg black hole using the EHT shadow-size measurements of Sgr A* and M87*. We summarize our findings in Section \ref{conclusion}. We have employed units where $8 \pi G = c = 1$; however, the tables reinstate these values.

\section{String-inspired Euler-Heisenberg Black holes}\label{solution}
In string theory, non-Maxwell NED fields naturally appear coupled with the dilaton field upon dimensional reduction. In one such framework, the four-dimensional, low-energy limit of an appropriate underlying string theory action in the presence of a dilaton field $\phi$ and a NED field $\mathcal{F}$ in the Einstein frame is given by \cite{Metsaev:1987ju,Bakopoulos:2024hah}
\begin{equation}
\begin{aligned}
\mathcal{S} =\frac{1}{16\pi} \int d^4x \sqrt{-g} \Big[\mathcal{R} - 2\nabla^{\mu}\phi\nabla_{\mu}\phi -e^{-2\phi}\mathcal{F}^2-\\f(\phi)\big(  2\alpha\mathcal{F}^{{i}}_{~{j}}\mathcal{F}^{{j}}_{~\gamma}\mathcal{F}^{\gamma}_{~\delta}\mathcal{F}^{\delta}_{~{i}}-\beta \mathcal{F}^4\big)\Big]~.
\label{theory}  
\end{aligned}
\end{equation}
It is a simplified version of the functional Einstein frame action, resulting from a non-diagonal reduction of the Gauss-Bonnet action \cite{Charmousis:2012dw}. For coupling $f(\phi)=0$ or $\alpha=\beta=0$, the exact solution of the theory is the well-known GMGHS black hole metric \cite{Gibbons:1987ps, Garfinkle:1990qj}. In the provided action \eqref{theory}, key components include $\mathcal{R}$ representing the Ricci scalar, $\mathcal{F}^2 \equiv \mathcal{F}_{\mu\nu}\mathcal{F}^{\mu\nu} \sim {\vec{E}}^{2} - {\vec{B}}^2$ as the Faraday scalar, and $\mathcal{F}^4 \equiv \mathcal{F}_{\mu\nu}\mathcal{F}^{\mu\nu}\mathcal{F}_{{i}{j}}\mathcal{F}^{{i}{j}}$, where $\mathcal{F}_{\mu\nu}$ denotes the field strength $\mathcal{F}_{\mu\nu} = \partial_{\mu}\mathcal{A}_{\nu} - \partial_{\nu}\mathcal{A}_{\mu}$, and $\alpha,\beta$ are coupling constants with dimensions of (length)$^2$, treated phenomenologically. The scalar field $\phi$ and the associated scalar function $f(\phi)$ are dimensionless. 
The field equations emanating from \eqref{theory} are of the following form \cite{Bakopoulos:2024hah}
\begin{equation}
\begin{aligned}\label{grav-eqs}
 G_{\mu\nu} &=2\partial_{\mu}\phi\partial_{\nu}\phi - g_{\mu\nu}\partial^{s}\phi\partial_{s}\phi + 2 e^{-2\phi} \Big( \mathcal{F}_{\mu}^{~ i}\mathcal{F}_{\nu i}-\frac{1}{4}g_{\mu\nu}\mathcal{F}^2\Big)\\
 &+f(\phi)\Big(8 \alpha~ \mathcal{F}_{\mu}^{~i}\mathcal{F}_{\nu}^{~j}\mathcal{F}_{i}^{~\eta}\mathcal{F}_{j\eta} -\alpha~g_{\mu\nu}\mathcal{F}^{i}_{~j}\mathcal{F}^{j}_{~\gamma}\mathcal{F}^{\gamma}_{~\delta}\mathcal{F}^{\delta}_{~i}\\
 &-4 \beta~\mathcal{F}_{\mu}^{~\xi}\mathcal{F}_{\nu\xi}\mathcal{F}^2 + \beta~\frac{1}{2}g_{\mu\nu} \mathcal{F}^4\Big),\\
4\square \phi &= -2e^{-2\phi}\mathcal{F}^2 +\frac{df(\phi)}{d\phi}\left(2\alpha\mathcal{F}^{i}_{~j}\mathcal{F}^{j}_{~\gamma}\mathcal{F}^{\gamma}_{~\delta}\mathcal{F}^{\delta}_{~i}-\beta \mathcal{F}^4\right),
\end{aligned}
\end{equation}
\begin{eqnarray}\label{eq3}
\partial_\mu\Big\{ \sqrt{-g} \Big[ && 4 \mathcal{F}^{\mu\nu}\left(2\beta f(\phi)\mathcal{F}^2-e^{-2\phi}\right),\nonumber\\\
    &&   -16 \alpha \mathcal{F}^{\mu}{}_\kappa \mathcal{F}^{\kappa}{}_\lambda \mathcal{F}^{\nu\lambda}\Big] \Big\}=0\,    
\end{eqnarray}       
The key difference from the GMGHS theory~\cite{Gibbons:1987ps, Garfinkle:1990qj} arises due to the presence of higher-order electromagnetic invariant terms $\mathcal{F}^4$ and $\mathcal{F}^{i}_{~j}\mathcal{F}^{j}_{\gamma}\mathcal{F}^{\gamma}_{~\delta}\mathcal{F}^{\delta}_{i}$, representing the Euler-Heisenberg NED field. 
With a nontrivial coupling function $f(\phi)$ \cite{Bakopoulos:2024hah},
\begin{equation} f(\phi) = - \frac{1}{2} \left(3 e^{-2 \phi } + 3 e^{2 \phi }+4\right),~\label{couplingfunction}
\end{equation}
and a purely \textit{magnetic} NED field $\mathcal{A}_{\mu}=(Q \cos\vartheta)\,\delta^{\varphi}_{\mu}$, the field equations are solved to get the string-inspired Euler-Heisenberg magnetically charged GMGHS black hole solution \cite{Bakopoulos:2024hah}:
\begin{equation}
\label{line-elem}
 {\rm d}s^2 = -F(r) {\rm d}t^2 + \frac{ {\rm d}r^2}{F(r)} + R(r)^2 ({\rm d}\vartheta^2+\sin^2\vartheta d\varphi^2),
\end{equation}
with
\begin{equation}
\begin{aligned}
F(r) &=1-\frac{2 M}{r}-\frac{2 (\alpha -\beta ) Q^4}{r^6 \left(1-\frac{Q^2}{M r}\right)^3},\\
R(r)^2 &=r^2 \left(1-\frac{Q^2}{Mr}\right).
\end{aligned}
\label{BR-r}
\end{equation}
Here, $M$ is the black hole's ADM mass, and $Q$ is the magnetic charge. On the other hand, for $\phi=0$ and $f(\phi)=1$, the field equations give the Einstein-Euler-Heisenberg black hole solution \cite{Yajima:2000kw}. 
The absence of the standard Reissner-Nordström (RN) term $Q^2/r^2$ in the metric function $F(r)$ is due to the dilaton coupling and the dominance of higher-order Euler-Heisenberg corrections. Unlike the RN solution, where the electromagnetic contribution appears at $O(1/r^2)$, the string-inspired NED terms introduce leading-order corrections at $O(1/r^6)$, significantly altering the strong-field behaviour.

Notably, the leading-order contribution of the NED field to the black hole metric (\ref{line-elem}) appears only at $\mathcal{O}(1/r^6)$ and is controlled by the relative strengths of $\alpha$ and $\beta$. The metric (\ref{line-elem}) is stable under linear and radial perturbations \cite{Bakopoulos:2024hah}. For $\alpha>\beta$ ($\alpha<\beta$), the NED terms act attractively (repulsively), and for $\alpha=\beta$, the NED field presented at the action level does not contribute to the black hole solution, reducing the metric (\ref{line-elem}) to the GMGHS solution \cite{Gibbons:1987ps, Garfinkle:1990qj}, and further to the Schwarzschild solution~\cite{Schwarzschild:1916abc} when $Q=0$. 

The metric function $R(r)$ measures the areal or curvature radius of a 2-sphere of coordinate radius $r$. We rewrite the metric (\ref{line-elem}) in a physical coordinate system where the radial coordinate $r$ represents the areal radius of a 2-sphere and also redefine the coordinates and parameters in units of the black hole mass $M$, to obtain the final expression of the black hole metric:
\begin{equation}
{\rm d}s^2=-A(x){\rm d}t^2+B(x){\rm d}x^2+C(x)({\rm d}\vartheta^2+\sin^2\vartheta d\varphi^2),
\label{5}
\end{equation}
where
\begin{equation}
\begin{aligned}
A(x)&=1-\frac{4}{q^2+\sqrt{q^4 + 4x^2}}-\frac{2\eta  q^4 }{x^6},\\
B(x)&=\frac{4x^2}{(q^4 + 4x^2)A(x)},~~~~ C(x)=x^2, \\
\eta&=\frac{\alpha-\beta}{M^2},\;\;\; q=\frac{Q}{M},\;\;\; x=\frac{r}{M}.
\end{aligned}
\label{6}
\end{equation}
In the limit $x\to\infty$, the string-inspired Euler-Heisenberg black hole metric (\ref{5}) converges to the magnetically charged RN black hole, therefore, from the weak-field test it is difficult to distinguish between the two metrics. Differences from the GMGHS and RN black holes are expected only in the strong-field regime. 

\begin{figure}
     \centering
     \includegraphics[scale=0.80]{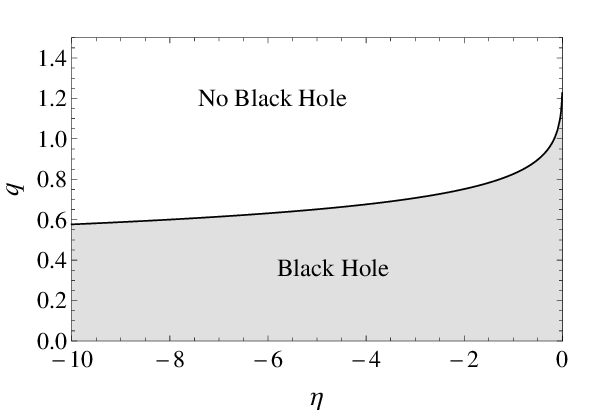}
     \caption{The parameter space [$\eta, q$] for the string-inspired Euler-Heisenberg black holes. The solid black line separates the black hole region from the no-black hole.}
     \label{parameterspace}
\end{figure}

\begin{figure}
    \includegraphics[scale=0.8]{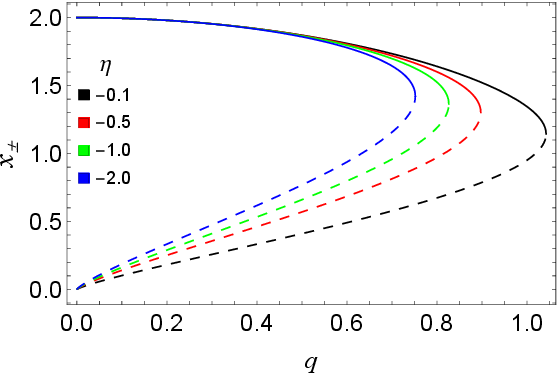}
    \caption{Radii of the event horizon ($x_+$) (solid lines) and Cauchy horizon ($x_-$) (dashed lines) for the string-inspired Euler-Heisenberg black holes with varying $q$ and $\eta$.}
    \label{figh}
\end{figure}

By analyzing the curvature invariant quantities $\mathcal{R}$ and $\mathcal{R}_{abcd}\mathcal{R}^{abcd}$,the divergence of the Kretschmann and Ricci scalars at $r=0$ confirms the presence of a curvature singularity. The leading-order behaviour $K\sim1/r^{20}$ indicates a more substantial divergence than in the Schwarzschild case $K\sim1/r^6$, highlighting the role of higher-order Euler-Heisenberg corrections in the deep strong-field regime (cf. Appendix \ref{append2}). The solution parameter space $(\eta, q)$ is illustrated in Fig.~\ref{parameterspace}. For $\eta\geq 0$, metric (\ref{5}) features black holes with a single horizon. In contrast, $\eta<0$ interpolates from a black hole with two distinct horizons to an extremal black hole with degenerate horizons and eventually to a naked singularity. In Fig.~\ref{parameterspace}, the solid line separating the black hole region from the naked singularity region corresponds to extremal string-inspired Euler-Heisenberg black holes with degenerate horizons. The gray (black hole) region corresponds to the values of the parameters $q\in(0,1.415)$ and $\eta\in(-\infty,0)$. As $q$ increases, the event horizon radius $x_{+}$ decreases slowly and that for the Cauchy horizon $x_{-}$ increases rapidly, merging with $x_{+}$ at certain values of $\eta$ (cf. Fig.~\ref{figh}). The black hole charge $ q$'s maximum value decreases with the magnitude of $|\eta|$.

\section{\label{lensing}Strong gravitational lensing by black hole}
 
\begin{figure*}
\begin{center}	
\begin{tabular}{p{9cm} p{9cm}}
    \includegraphics[scale=.9]{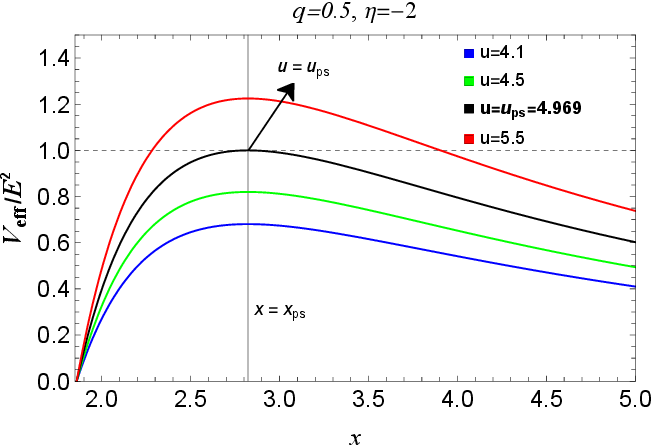}&\hspace{10mm}
    \includegraphics[scale=.78]{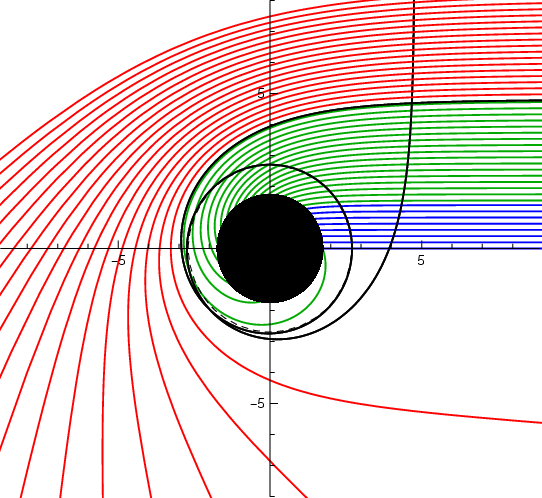}
\end{tabular}    
\caption{ Radial effective potential $V_{\rm eff}$ for string-inspired Euler-Heisenberg black holes with $q=0.5$ and $\eta=-2$ and with varying photon's impact parameter $u$ are shown. Photons with $u_{\rm ps}$, shown by the black line, follow an unstable circular orbit around the black hole (left).  Various light orbits around black holes, with $q=0.5$ and $\eta=-2$, are shown in polar coordinates ($r,\phi$). The black line corresponds to the value of  $u\approx u_{\rm ps}$. A circular disk shows a black hole, and the unstable photon orbit is a dashed black circle (right).}\label{fig:veff}
\end{center}      
\end{figure*}

Null geodesic equations for photons are discussed in detail in the Appendix~\ref{Apd-1}. Here, we consider that the photons follow the null geodesics of the background metric $g_{\mu\nu}$. 
Considering the propagation of photons on the equatorial plane, the radial motion is defined by 
\begin{equation}\label{Veff}
A(x)B(x)\left(\frac{dx}{d\tau}\right)^2 +  \frac{A(x)}{C(x)}\mathcal{L}^2=\mathcal{E}^2,
\end{equation}
with the radial effective potential
\begin{equation}
V_{\rm eff}= \frac{A(x)}{C(x)}\mathcal{L}^2.   
\end{equation}
Here, $\mathcal{E}=-p_\mu\xi_t^\mu$ and $\mathcal{L}={p_\mu\xi}_\phi^\mu$ are the conserved energy and angular momentum of the photon, such that $\xi_t^\mu$ and $\xi_\phi^\mu$ are killing vectors of spacetime, associated with time translation symmetry and axial rotational symmetry, respectively. The radial velocity vanishes $\dot{x}=0$ at the point of closest approach $x_0$, that can be used to define the impact parameter $u$ as follows \cite{Bozza:2002zj}
\begin{equation}\label{impact}
u\equiv \left|\frac{\mathcal{L}}{\mathcal{E}}\right| =\sqrt{\frac{C(x_0)}{A(x_0)}}.
\end{equation}
Photons can propagate only in the region where $V_{\text{eff}}/\mathcal{E}^2 \leqslant 1$. Additionally, one can define an unstable (or stable) circular orbit that satisfies $V_{\text{eff}}(x_{\text{ps}}) =\mathcal{E}^2,\,  V_{\text{eff}}'(x_{\text{ps}}) = 0$ and $V_{\text{eff}}''(x_{\text{ps}}) > 0$ (or $V_{\text{eff}}''(x_{\text{ps}}) < 0$). For string-inspired Euler-Heisenberg black holes, we observe that $V''_{\text{eff}}(x_{\text{ps}}) > 0$, indicating unstable photon circular orbits (cf. Fig. \ref{fig:veff}). Photons move along these unstable circular orbits around a black hole without falling into it or escaping to infinity. However, even a small radial perturbation can cause the photons to fall into the black hole or escape to infinity. 

A direct consequence of the underlying spherical symmetry is that the generators of two spatial Killing vectors fix the orbital plane. In contrast, the ratio of generators of $\xi_\phi^\mu$ and $\xi_t^\mu$ solely determines the radial motion in a plane. Therefore, when a photon travels from an emitter to the observer, it moves along a spatially-planar orbit centred around the black hole. Once the plane of a photon orbit is fixed, the impact parameter $u$ solely determines the radial motion, controlling the distance of the closest approach and the image positions on the observer's screen. Even at $\vartheta\neq \pi/2$ plane, photons follow circular orbits of constant radii that altogether form a photon sphere around the black hole\footnote{For motion at $\vartheta\neq \pi/2$, geodesics admit a fourth constant of the motion, the Carter constant $\mathcal{K}$. Photons moving in the $\vartheta= \pi/2$ plane have $\mathcal{K}=0$, resulting in a (constant$-\vartheta$) equatorial orbit purely in $\varphi$, and those with $\mathcal{L}=0$ and $\mathcal{K}\neq 0$ correspond to meridional (constant$-\varphi$) orbits purely in $\vartheta$. Although photons with $\mathcal{K}\neq 0, \mathcal{L}\neq 0$ exhibit both $\vartheta$ and $\varphi$ motion, the spatial projections of all of these orbits are always planar.}. The gravitationally-lensed projection of this photon sphere on the observer's image plane results in the formation of the black hole shadow. For a static and spherically symmetric black hole, the shadow is always circularly symmetric and independent of the observer's viewing angle. 

The radius of the photon sphere $x_{\rm ps}$ is the largest positive root of the following equation \cite{Claudel:2000yi,Virbhadra:2002ju}:
\begin{equation}
\frac{A'(x)}{A(x)}=\frac{C'(x)}{C(x)},\label{root}
\end{equation}
The photon sphere radii $x_{\rm ps}$ decreases with the parameters $q$ and $\eta$. Although these constant-radii photon orbits are unobservable, only nearly bound unstable orbits yield observable images.

For light rays at the equatorial plane, the change in azimuthal angle from the source to the observer defines the light deflection angle, mathematically expressed as: \cite{Virbhadra:1999nm} as
\begin{equation}
\alpha_D(x_0)=I(x_0)-\pi,
\label{24}
\end{equation}
where
\begin{equation}
I(x_0)=2\int_{x_0}^{\rm \infty}\frac{d\varphi}{dx}dx,
\label{25}
\end{equation}
with
\begin{equation}
\frac{d\varphi}{dx}=\frac{\sqrt{B(x)}}{\sqrt{C(x)}\sqrt{\frac{A(x_0)C(x)}{A(x)C(x_0)}-1}}.
\label{26}
\end{equation}
Without a black hole, photons travel along a straight-line trajectory, resulting in a deflection angle of zero by definition. However, in the presence of a black hole, as the distance of the closest approach decreases, the deflection angle increases monotonically, such that as $x_0 \to x_{\rm ps}$ or $u \to u_{\rm ps}$, light ray makes multiple loops around the black hole before reaching the observer and the deflection angle becomes unboundedly large \cite{Virbhadra:1999nm}. This phenomenon leads to the formation of infinite or relativistic images of the light source.

We define a new variable $z=1-x_0/x$ \cite{Tsukamoto:2016jzh,Tsukamoto:2017fxq}, to rewrite the Eq.~(\ref{25}) as 
\begin{equation}
I(x_0)=\int_{0}^{1}{R(z,x_0)f(z,x_0)}dz.
\label{28}
\end{equation}
with
\begin{eqnarray}
R\left(z,x_0\right)&=&\frac{2x_0}{\left(1-z\right)^2}\frac{\sqrt{B(x)A(x)C\left(x_0\right)}}{C\left(x\right)},\label{29}\\
f(z,x_0)&=&\frac{1}{\sqrt{A(x_0)-A(x)\frac{C(x_0)}{C(x)}}}.\label{30}
\end{eqnarray}
The motivation is to identify the diverging term; $R(z,x_0)$ is regular, but the function $f(z,x_0)$ diverges when $z=0$. Therefore, to avoid the divergence, we do a Taylor series expansion of the function $f(z,x_0)$ at $z=0$ to approximate the function $f(z,x_0)\approx f_0(z,x_0)$ \cite{Tsukamoto:2017fxq}
\begin{equation}
f_0(z,x_0)=\frac{1}{\sqrt{\Gamma(x_0)z+\gamma(x_0)z^2}}.
\label{31}
\end{equation}
where,
\begin{eqnarray}
 &\Gamma =\frac{1-A_0}{C_0 A'_0} (C'_0 A_0-C_0 A'_0
)\label{Gamma1} \\
 &\gamma = \frac{\left( 1-A_0 \right)^2}{2C_0^2 {A'_0}^3}[
2C_0 C'_0 {A'_0}^2 + \left(C_0 C''_0-2{C'_0}^2 \right)A_0 A'_0 \nonumber
\\
&-C_0 C'_0 A_0
A''_0 ]. \label{gamma2}
\end{eqnarray}

Our analysis confirms that for extremal configurations ($x_+=x_{-}$), the coefficient $\Gamma$ in Eq.~(\ref{Gamma1}) can indeed vanish at critical impact parameters, modifying the deflection angle's logarithmic divergence to a stronger power-law form as discussed in \cite{Khoo:2016xqv,Ulbricht:2015vwa,Chen:2024oyv}. Notably, we find $\Gamma$ may also vanish for certain non-extremal cases when the metric functions satisfy $C'(x_0) A(x_0)-C(x_0) A'(x_0)$, even with distinct horizons.
The integral can be divided into two parts as
\begin{equation}
I(x_0)=I_D(x_0)+I_R(x_0),
\label{34}
\end{equation}
where $I_D(x_0)$ is the divergent part described as
\begin{equation}
I_D(x_0)=\int_{0}^{1}{R(0,x_{ps})f_0(z,x_0)}dz,
\label{35}
\end{equation}
and $I_R(x_0)$ is the regular part
\begin{equation}
I_R(x_0)=\int_{0}^{1}\left(R(z,x_0)f(z,x_0)-R(0,x_{ps})f_0(z,x_0) \right)dz.\label{36}
\end{equation}
By solving both the integrals \eqref{35} and \eqref{36}, the deflection angle, as a function of impact parameter $u$, can be written as \cite{Bozza:2002zj,Bozza:2003cp}
\begin{equation}
\alpha_D(u)=-\bar{a}\log\left(\frac{u}{u_{\rm ps}}-1\right)+\bar{b}+O(u-u_{\rm ps}).
\label{37}
\end{equation}
where $\bar{a}$ and $\bar{b}$ are the strong lensing coefficients described as
\begin{eqnarray}
\bar{a}&=&\frac{R(0,x_{\rm ps})}{2\sqrt{\gamma\left(x_{\rm ps}\right)}},\label{38}\\
\bar{b}&=&-\pi+I_R(x_{\rm ps})+\bar{a}\log\left(\frac{2\gamma(x_{\rm ps})}{A(x_{\rm ps})}\right),\label{39}
\end{eqnarray}
with
\begin{equation}
   \gamma(x_{ps})=\frac{C_{ps}\left( 1-A_{ps} \right) ^2\left( A_{ps}C_{ps}^{''}-A_{ps}^{''}C_{ps} \right)}{2A_{ps}^{2}C_{ps}^{\prime2}} \label{gamma}.
\end{equation}

\begin{figure*}[!]
\begin{centering}
\begin{tabular}{p{9cm} p{9cm}}
    \includegraphics[scale=0.85]{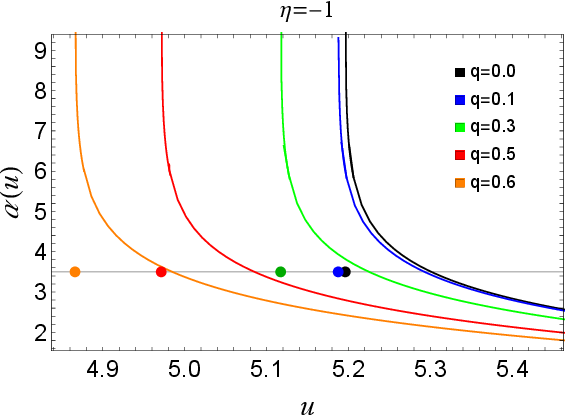}&
    \includegraphics[scale=0.85]{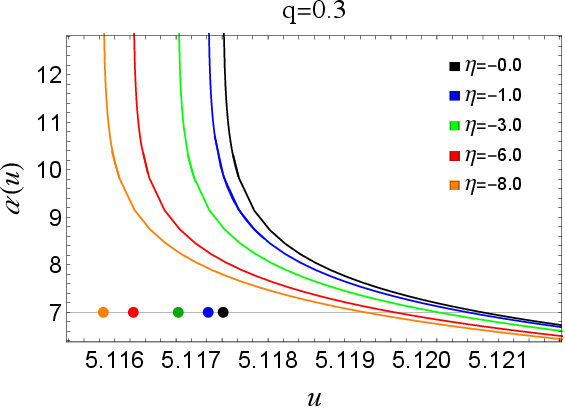}   
\end{tabular}
\end{centering}
\caption{The effects of independently varying the black hole charge $q$ (right) and coupling constant $\eta$ (left) on the light deflection angle are shown. The deflection angle, $\alpha(u)$, increases monotonically with decreasing impact parameters. Critical values $u_{\rm ps}$ are indicated by coloured points on the horizontal axis, where $\alpha(u)$ diverges logarithmically. For comparison, the deflection angle for GMGHS black holes ($\eta=0$) is shown in the right figure.}
\label{def} 
\end{figure*}

This determines the relationship between the strong gravitational lensing coefficients, deflection angle, and string-inspired Euler-Heisenberg black hole parameters. The deflection coefficient $\bar{a}$ increases with parameters $q$ and $\eta$, while the deflection coefficient $\bar{b}$ decreases with these parameters. The critical impact parameter decreases as $q$ increases and $\eta$ decreases (see Table \ref{tab1}). Note that when $q$ and $\eta$ vanish, the string-inspired Euler-Heisenberg black hole becomes a Schwarzschild black hole, resulting in $\bar{a}=1$ and $\bar{b}=-0.40023$. Figure \ref{def} illustrates that the deflection angle diverges at specific impact parameter values $u=u_{\rm ps}$ for different $q$. The light deflection angle for the string-inspired Euler-Heisenberg black hole gradually decreases as the parameter $q$ increases for the fixed value of parameter $\eta$ (cf. Fig. \ref{def}). Note that the results presented are valid only in the strong deflection limit; for $u\gg u_{\rm ps}$, the strong deflection limit is not a valid approximation.

\begin{table}[h!]
 \begin{centering}	
	\begin{tabular}{p{1cm} p{1.5cm} p{1.5cm} p{1.5cm} p{1.5cm}  }
 \vspace{1mm}\\
\hline\hline
\vspace{0.01mm}\\
{$q$ } & {$\eta$}& {$\Bar{a}$} & {$\Bar{b}$}  & {$u_{\rm ps}/R_s$}  \\ \hline\hline
\vspace{0.01mm}\\
\multirow{1}{*}{$0.0$} & $0.0$ &  $1.00$ & -0.40023 & $5.19615$ \\
\hline
\multirow{5}{*}{0.1}&0.0&1.0011 & -0.400001 & 5.18748 \\
&-0.1 & 1.00111 & -0.400002 & 5.18748 \\
 &  -0.5 &1.00112 & -0.400004 & 5.18748 \\
  &  -1& 1.00112 & -0.400006 & 5.18748 \\
  &  -2& 1.00112 & -0.400011 & 5.18748\\            
\hline 
\multirow{5}{*}{0.3}&0.0 &1.01026 & -0.398122 & 5.11741 \\
&  -0.1  & 1.01028 & -0.39817 & 5.11739 \\
 &  -0.5 &  1.01039 & -0.398362 & 5.11732 \\
 &  -1 &  1.01052 & -0.398601 & 5.11722\\
  &  -2& 1.01078 & -0.399082 & 5.11702 \\     
\hline
\multirow{5}{*}{0.5}&0.0 &1.0299 & -0.39408 & 4.97324 \\
&  -0.1  & 1.03018 & -0.394593 & 4.97305 \\
 &  -0.5 & 1.03128 & -0.396662 & 4.9723 \\
&  -1 & 1.03267 & -0.399283 & 4.97135 \\
 &  -2& 1.03549 & -0.404648 & 4.96945 \\
\hline\hline
\end{tabular}
\end{centering}
\caption{ Strong lensing coefficients $\bar{a}$ and $\bar{b}$ and the critical impact parameter $u_{\rm ps}$ with varying black hole parameters. Note that $\eta=0$ and $q\neq0$ correspond to GMGHS black holes and $u_{ps}$ is in the units of Schwarzschild radius
$R_s = 2G M/c^2$.}\label{tab1}
\end{table}

\section{Strong gravitational lensing effect for supermassive black holes}\label{observation}
\subsection{Strong lensing Observables}\label{character}
Considering both the source and observer far from the black hole, the following lens equation relates the source position with their image positions \cite{Virbhadra:1999nm,Bozza:2001xd}
\begin{equation}
\psi=\theta-\frac{D_{\rm LS}}{D_{\rm OS}}\Delta\alpha_n,
\label{40}
\end{equation}
where $D_{\rm LS}$ is the distance between the lens (black hole) and the source, and $D_{\rm OS}$  is the distance between the observer and the light source. Also, the angular positions of the source and image are $\psi$ and $\theta$, respectively, and $\Delta\alpha_n=\alpha(\theta)-2n\pi$ is the offset of deflection, where $n$ is the integer representing the order of image. Substituting the expression of the deflection angle, the position of the $n$th relativistic image is determined by \cite{Bozza:2002zj}.
\begin{equation}
\theta_n=\theta_n^0+\frac{u_{\rm ps}e_n(\psi-\theta_n^0)D_{\rm OS}}{\bar{a}D_{\rm LS}D_{\rm OL}}.
\label{43}
\end{equation}
where
\begin{equation}
e_n=\exp\left(\frac{\bar{b}-2n\pi}{\bar{a}}\right).
\label{42}
\end{equation}
Here, $\theta_n^0$ corresponds to the image position for deflection angle $\alpha=2n\pi$. According to Eq.~(\ref{43}), when $\psi=\theta_n^0$, i.e., source and its image are on the same side of the black hole and share the angular position, $\theta_n=\theta_n^0$. However, to locate the $n$th image on the opposite side of the source, this condition is extended by substituting $\psi$ with -$\psi$.
The magnification of the $n$th relativistic image can be defined as \cite{Virbhadra:1998dy,Bozza:2002zj,Virbhadra:2007kw}
\begin{equation}
\mu_n=\left.\left(\frac{\psi}{\theta}\frac{d\psi}{d\theta}\right)^{-1}\right|_{\theta_n^0}=\frac{{u_{\rm ps}}^2e_n(1+e_n)D_{\rm OS}}{\bar{a}\psi D_{\rm LS}{D_{\rm OL}}^2},
\label{flux}
\end{equation}
where the first relativistic image is the brightest, and the magnification decreases exponentially with increasing $n$. The relativistic images are faint since the magnification is inversely proportional to the square of the distance between the observer and the lens. 

When a black hole is perfectly aligned ($\psi\approx0$) with the observer and the source and is equidistant from them, the angular position of the image simplifies to \cite{Bozza:2004kq}
\begin{equation}
\theta_n^E=\left(1-\frac{u_{\rm ps}e_nD_{\rm OS}}{\bar{a}D_{\rm LS}D_{\rm OL}}\right)\ \theta_n^0,
\label{44}
\end{equation}
where $\theta_n^E$ is the Einstein ring \cite{Einstein:1936llh}.
Now assuming that the distance $D_{\rm OL}$  is greater than the impact parameter i.e. $u_{\rm ps}(D_{\rm OL}\gg u_{\rm ps}$, then (\ref{44}) becomes 
\begin{equation}
\theta_n^E\approx\frac{u_{\rm ps}(1+e_n)}{D_{\rm OL}}.
\label{51}
\end{equation}
At $n=1$, the outermost ring is formed, and the radius of the ring decreases with increases in the value of $n$. For Einstein rings $\psi= 0$, magnification is maximum.

We now calculate three lensing observables to analyze the strong gravitational lensing phenomena. Here, we consider the outermost image $\theta_1$ and all the other inner-packed images as $\theta_\infty$  \cite{Bozza:2002zj}.
The angular position of the asymptotic relativistic image $\theta_\infty$,
\begin{equation}
\theta_\infty=\frac{u_{\rm ps}}{D_{\rm OL}},
\label{46}
\end{equation}
the angular separation $s$ between the outermost and all the other images is
\begin{equation}
s\equiv\theta_1-\theta_\infty=\theta_\infty \exp\left(\frac{\bar{b}-2\pi}{\bar{a}}\right),
\label{s}
\end{equation}
the flux ratio of the outermost image $\theta_1$ and the remaining relativistic images at $\theta_\infty$ is
\begin{equation}
r=\frac{\mu_1}{\sum_{n=2}^{\rm \infty}\mu_n}=2.5\log_{10}\left[\exp\left(\frac{2\pi}{\bar{a}}\right)\right],
\label{rmag}
\end{equation}
The flux ratio is independent of the distance between the lens and the observer $D_{\rm OL}$. One can obtain the coefficients $\bar{a}$, $\bar{b}$ and the critical impact parameter $u_{\rm ps}$, using the three lensing observables defined above to analyze the properties specific to string-inspired Euler-Heisenberg black holes. The strong-field features of string-inspired Euler-Heisenberg black holes can be understood by comparing astronomical observations with the analysed data.

\subsection{\label{sbh}Strong Lensing effect for the supermassive black holes M87* and SgrA*}

We model EHT targets, supermassive black holes M87* and Sgr A*, as string-inspired Euler-Heisenberg black hole models to estimate their lensing observables and compare our results with those for Schwarzschild and GMGHS black holes. Based on the latest astronomical observation data, the estimated mass of M87* is $\left(6.5\pm0.7\right)\times{10}^9M_\odot$, and its distance is $d=16.8$ MPc \cite{EventHorizonTelescope:2019ggy}. The estimated mass of SgrA* is $4_{-0.6}^{+1.1}\times{10}^6M_\odot$, and its distance is $d=8.15\pm0.15$ KPc \cite{EventHorizonTelescope:2022xqj}.

The radii of the outermost relativistic Einstein rings of Sgr A*, M87*, NGC 4649, and NGC 1332 are $25.9289~\mu as$, $19.48077~\mu as$, $14.14604~\mu as$ and $4.40565~\mu as$ respectively, at the value of the parameter $\eta=-2$ and $q=0.3$.

We compare the relativistic image positions $\theta_\infty$ and lensing observables, $s$ and $r_{mag}$, for string-inspired Euler-Heisenberg black holes with those from the Schwarzschild black hole and summarize the results in Table \ref{Table2a}. The deviation in $\theta_\infty$ and $s$ between the string-inspired Euler-Heisenberg black holes and their GR counterpart is not very large (around 1$\mu$as). As shown in Table \ref{Table2a}, $\theta_\infty$ and $r_{\rm mag}$ decrease as the parameters $q$ and $\eta$ increase in magnitude. Similarly, the image separation $s$ increases with $q$ and $\eta$.
For M87* and Sgr A*, the angular positions of the relativistic images vary within the ranges of $19.78\mu as\geq \theta_\infty\geq 9.12\mu as$ for M87* and $26.33\mu as\geq \theta_\infty\geq 11.52\mu as$ for Sgr A*.  Furthermore, the separation between the outermost image and the shadow boundary $s$ ranges as 32.95-39.45 $nas$ for Sgr A* while 24.75-40.2 $nas$ for M87* respectively.

\begin{figure*}[!]
	\begin{centering}
		\begin{tabular}{p{9cm} p{9cm}}
		    \includegraphics[scale=0.85]{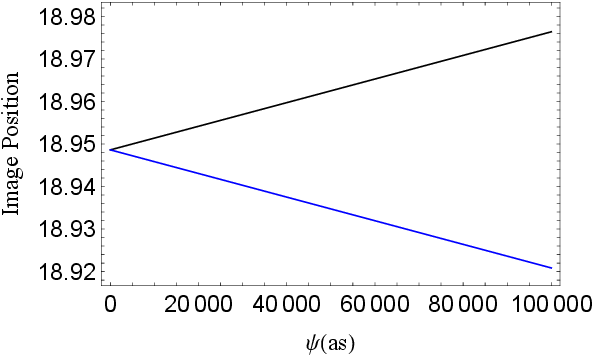}&
			\includegraphics[scale=0.92]{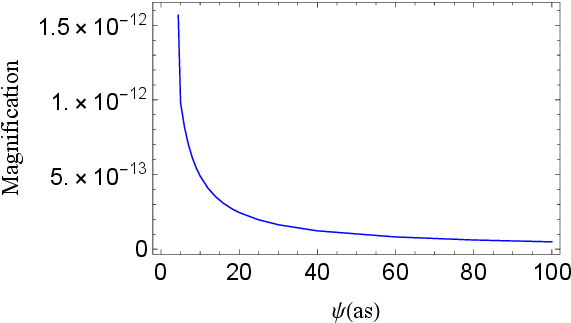}
    \label{tab:table1}
    \end{tabular}
	\end{centering}
	\caption{Comparison between the position of first primary $\theta_{1p}$ (solid black line) and secondary image $\theta_{1s}$ (solid blue line) for different values of the position of the source $\psi$ for M87*(Left). The absolute magnification of the first primary image ($\mu_{1p}$) is plotted against source position $\psi$ (Right). }\label{comparison}		
\end{figure*}

\begin{figure*}
    \centering
    \begin{tabular}{p{8cm} p{9cm}}
    \includegraphics[scale=0.8]{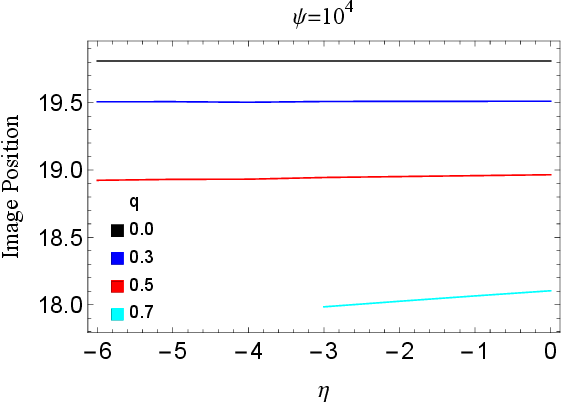}&
    \includegraphics[scale=0.8]{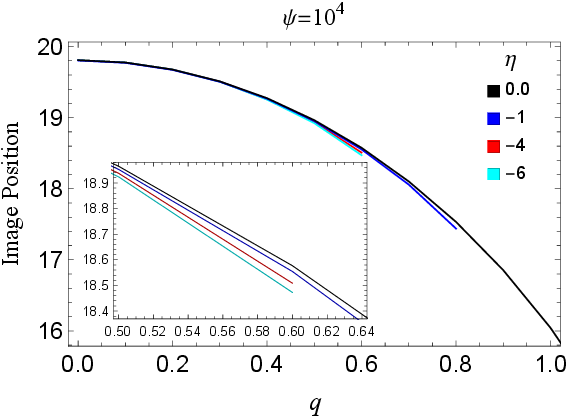}
    \end{tabular}
    \caption{Variation of the position of the primary image $\theta_{1p}$ (solid line) and secondary image $\theta_{1s}$ (dashed line) in $\mu as$ as a function of $\eta$ and $\psi$, shown for different values of the parameter $q$ for M87*. }
    \label{imageposition}
    
\end{figure*}

With the help of Eq.~(\ref{40}) and (\ref{43}), we compute the angular positions of $n$th-order relativistic images, which are produced when the photons make $n$ loops around the black hole before reaching the observer. Taking $d=D_{\rm LS}/D_{\rm OS}=1/2$ for black holes Sgr A * and M87*, we computed the angular positions of the secondary images for different values of the parameters $q$ and $\eta$ at different source positions $\psi$. Note that the negative sign of the secondary image represents the images on the opposite side of the black hole to the source. From the results presented in Tables \ref{tablesgr} and \ref{tablem87}, we confirm that for Euler-Heisenberg black holes, the angular positions of the images are smaller compared to their GR counterparts, with only a weak dependence on $\psi$. Moreover, relativistic images in the Euler-Heisenberg black hole spacetime appear closer to the black hole than those in the Schwarzschild black hole case. The angular positions $|\theta_{1p}| > |\theta_{1s}|$ for higher values of $\psi$ (cf. Fig.~\ref{comparison}), however, for small values of $\psi$ the values are extremely close. The same is true for any pair of higher-order relativistic images. The deviation of primary and secondary images is too small to be observed with the EHT telescope. We have also plotted the primary image position for different values of the parameters $\eta$ and $q$ for the constant value of the source angular position $\psi$ (c.f. Fig~\ref{imageposition}). The relative magnifications of the Euler-Heisenberg black holes and their ratio with that of Schwarzschild black hole for the secondary images for SgrA* and M87* are also summarized in Tables \ref{tablesgr} and \ref{tablem87}, respectively. We noticed that higher-order images are significantly demagnified, but the ratio of brightness of relativistic images increases for higher-order images for higher magnitude of parameters $\eta$ and $q$.

\begin{table}[h!]

 \begin{centering}	
	\begin{tabular}{p{0.6cm} p{.8cm} p{1.2cm} p{1.5cm} p{1.2cm}  p{1.5cm} p{1.2cm} }
 \vspace{1mm}\\
\hline\hline
\multicolumn{2}{c}{}&
\multicolumn{2}{c}{Sgr A*}&
\multicolumn{2}{c}{M87*}\\
{$q$ } & {$\eta$}& {$\theta_{\rm \infty}$($\mu$as)} & {$s$ ($\mu$as)}  & {$\theta_{\rm \infty}$($\mu$as)}  & {$s$ ($\mu$as) } & {$r_{\rm mag}$}  \\ \hline
\hline

\multirow{1}{*}{0.0}  &0.0 & 26.33 & 0.03295 &19.78 & 0.024757 & 6.8219 \\
\hline
\multirow{5}{*}{0.1}  & 0.0&26.28 &0.03314 & 19.7490 & 0.024906 & 6.81429 \\
&-0.1 &26.2859 & 0.03315 &19.749 & 0.0249061 &  6.81429 \\
& -1.0 &  26.2859 & 0.0331504 &19.749 & 0.0249064 &  6.81427 \\ 
& -4.5 & 26.2859 & 0.0331519& 19.749 & 0.0249075 &6.8142 \\
&-9.5 &  26.2858 & 0.0331541 & 19.7489 & 0.0249092 & 6.81411 \\   
\hline 

\multirow{5}{*}{0.3} &0.0&  25.9309 & 0.0348038 & 18.757 & 0.02991 & 6.75261 \\ 
& -0.1 &25.9308 & 0.034808 &18.7562 & 0.02996 & 6.75244 \\
& -1.0 &25.9299 & 0.0348456 &18.7462 & 0.03048 &6.75086\\
& -2.5 & 25.9284 & 0.0349083 &18.7292 & 0.03137 &6.74824 \\ 
&-4.5 & 25.9264 & 0.0349924 & 18.7061 & 0.03263 & 6.74473  \\ 
\hline

\multirow{5}{*}{0.5}&0.0& 25.2003& 0.0385218 & 18.028 & 0.03437 & 6.62381 \\
& -0.1 &25.1994 & 0.0385676 &18.0246 & 0.03462 &  6.62204 \\ 
& -0.5 & 25.1956 & 0.0387519 &18.0074 & 0.03568 & 6.61497 \\
& -1.0 & 25.1908 & 0.0389847 &17.9855 & 0.03709 & 6.60607 \\
&-2.0 &25.1811 & 0.0394587 &17.9399 &  0.0402 &6.58805 \\            
\hline\hline  
\end{tabular}
\end{centering}
\caption{Estimation of strong lensing observables for the string-inspired Euler-Heisenberg black holes. We compare these observables with those for Schwarzschild black holes. The comparison considers the supermassive black holes Sgr A* and M87* as the lens for different values of the parameters $q$ and $\eta$. Note that $q\neq0$ and $\eta=0$ corresponds to GMGHS black holes. }\label{Table2a}
    \end{table}
\begingroup
\begin{table*}
	\caption{Image positions and magnifications of first and second order secondary images due to lensing by Sgr A* with $d=D_{\rm LS}/D_{\rm OS}=1/2$. Schwarzschild and Euler-Heisenberg black holes prediction with the parameters $q$ and $\eta$ for different values of angular source position $\psi$. {\bf (a)} All angles are in $\mu${\em as}. {\bf (b)} We have used $M_{\text{Sgr A*}}= 4.3\times 10^6	M\odot$, $D_{\rm OL}= 8.3 Mpc$ {\bf (c)} Note that $q\neq0$ and $\eta=0$ corresponds to GMGHS black holes and $\Delta\theta_{ns}=\theta_{\rm ns, EH}-\theta_{\rm ns, Schw}$.{\bf (d)} Negative sign of secondary image represent the image is on the opposite side of the source.}\label{tablesgr}
	\begin{ruledtabular}
		\begin{tabular}{l cccc cccc ccc}
			
   
$\psi\text{(as)}$&{$q$ } & {$\eta$}& $\theta_{1s,{\rm EH}} $ &$\theta_{2s,{\rm EH}} $&$\mu_{1s,{\rm EH}}$  &$\mu_{2s,{\rm EH}}$  &$\Delta\theta_{1s}$&$\Delta\theta_{2s}$& $\frac{\mu_{\rm 1s,EH}}{\mu_{\rm 1s,Schw}}$&$\frac{\mu_{\rm 2s,EH}}{\mu_{\rm 2s,Schw}}$\\ 
			\hline\hline
		\vspace{1mm}
   \multirow{4}{*}{10}  &0.1&0.0& -26.3191&-26.286 &$-8.450\times10^{-13}$&$-1.587\times10^{-15}$& 0.0437&0.041&1.003&1.01\\
   &0.2&-1.0&-26.1871&-26.1534&$-8.534\times10^{-13}$&$-1.637\times10^{-15}$&0.1757&0.1765&1.013&1.04\\
   &0.3&-1.5&-25.9643&-25.9295&$-8.685\times10^{-13}$&$-1.730\times10^{-15}$&0.3985&0.4004&1.031&1.101\\
   &0.5&-2.0&-25.2206&-25.1812&$-9.318\times10^{-13}$&$-2.155\times10^{-15}$&1.0123&1.1422&1.106&1.37\\
\hline
 \multirow{4}{*}{$10^3$} &0.1 &0.0 & -26.3187 &-26.286&$-8.450\times10^{-16}$&$-1.587\times10^{-18}$&0.043& 0.0439 &1.003&1.01\\
   &0.2&-1.0&-26.1868&-26.1534&$-8.534\times10^{-16}$&$-1.637\times10^{-18}$&0.1757&0.1769&1.013&1.04\\
   &0.3&-1.5&-25.9639&-25.9295&$-8.685\times10^{-16}$&$-1.730\times10^{-18}$&0.3986&0.4004&1.031&1.101\\
   &0.5&-2.0&-25.2202&-25.1812&$-9.318\times10^{-16}$&$-2.155\times10^{-18}$&1.0123&1.1487&1.106&1.37\\
\hline
 \multirow{4}{*}{$10^5$} &0.1 &0.0 & -26.287 &-26.2859&$-8.450\times10^{-18}$&$-1.587\times10^{-20}$&0.0439&0.044&1.003&1.01 \\
   &0.2&-1.0&-26.1545&-26.1534&$-8.534\times10^{-18}$&$-1.637\times10^{-20}$&0.1764&0.1765&1.013&1.04\\
   &0.3&-1.5&-25.9308&-25.9294&$-8.685\times10^{-18}$&$-1.730\times10^{-20}$&0.4015&0.4005&1.031&1.101\\
   &0.5&-2.0&-25.1837&-25.1812&$-9.318\times10^{-18}$&$-2.155\times10^{-20}$&1.1472&1.1487&1.106&1.37\\
\hline
		\end{tabular}
	\end{ruledtabular}
\end{table*}
\endgroup

\begingroup
\begin{table*}
	\caption{Image positions and magnifications of first and second order secondary images due to lensing by M87* with $d=D_{\rm LS}/D_{\rm OS}=1/2$. Schwarzschild and Euler-Heisenberg black holes prediction with the parameters $q$ and $\eta$ for different values of angular source position $\psi$. {\bf (a)} All angles are in $\mu${\em as}. {\bf (b)} We have used $M_{\text{M87*}}=  6.5 \times 10^9 M\odot$, $D_{\rm OL}=  16.8  \times 10^6 Mpc$.{\bf (c)}  Note that, $q\neq0$ and $\eta=0$ corresponds to GMGHS black holes and $\Delta\theta_{ns}=\theta_{\rm ns,EH}-\theta_{\rm ns,Schw}$.{\bf (d)} Negative sign of secondary image represent the image is on the opposite side of the source. }\label{tablem87}
	\begin{ruledtabular}
		\begin{tabular}{l cccc cccc ccc}
			
   
$\psi\text{(as)}$&{$q$ } & {$\eta$}& $\theta_{1s,{\rm EH}} $ &$\theta_{2s,{\rm EH}} $&$\mu_{1s,{\rm EH}}$  &$\mu_{2s,{\rm EH}}$  &$\Delta\theta_{1s}$&$\Delta\theta_{2s}$& $\frac{\mu_{\rm 1s,EH}}{\mu_{\rm 1s,Schw}}$&$\frac{\mu_{\rm 2s,EH}}{\mu_{\rm 2s,Schw}}$\\ 
			\hline\hline
		\vspace{1mm}
   \multirow{4}{*}{10}  &0.1&0.0&-19.7739 &-19.7490&$-4.770\times10^{-13}$&$-8.958\times10^{-16}$&0.033&0.3331&1.003& 1.01 \\
   &0.2&-1.0&-19.6747&-19.6494&$-4.817\times10^{-13}$&$-9.243\times10^{-16}$&0.132&0.1327&1.013&1.042\\
   &0.3&-1.5&-19.5073&-19.4811&$-4.902\times10^{-13}$&$-9.769\times10^{-16}$&0.3257&0.301&1.03&1.101\\
   &0.5&-2.0&-18.9486&-18.9190&$-5.260\times10^{-13}$&$-12.16\times10^{-16}$&0.8878&0.8631&1.106&1.371\\
\hline
 \multirow{4}{*}{$10^3$} &0.1 &0.0 &-19.7736&-19.7490&$-4.770\times10^{-16}$&$-8.958\times10^{-19}$&0.0329&0.033 &1.003&1.01 \\
   &0.2&-1.0&-19.6745&-19.6494&$-4.817\times10^{-16}$&$-9.243\times10^{-19}$&0.132&0.1327&1.013&1.042\\
   &0.3&-1.5&-19.5071&-19.4811&$-4.902\times10^{-16}$&$-9.769\times10^{-19}$&0.2994&0.301&1.03&1.101\\
   &0.5&-2.0&-18.9483&-18.9190&$-5.260\times10^{-16}$&$-12.16\times10^{-19}$&0.8582&0.8631&1.106&1.371\\
\hline
 \multirow{4}{*}{$10^5$} &0.1 &0.0 &-19.7498&-19.7490&$-4.770\times10^{-18}$&$-8.958\times10^{-21}$&0.033&0.33&1.003&1.01  \\
   &0.2&-1.0&-19.6503&-19.6494&$-4.817\times10^{-18}$&$-9.243\times10^{-21}$&0.1325&0.1326&1.013&1.042\\
   &0.3&-1.5&-19.4822&-19.4811&$-4.902\times10^{-18}$&$-9.769\times10^{-21}$&0.3002&0.3009&1.03&1.101\\
   &0.5&-2.0&-18.9208&-18.9189&$-5.260\times10^{-18}$&$-12.16\times10^{-21}$&0.8345&0.8612&1.106&1.371\\
\hline
		\end{tabular}
	\end{ruledtabular}
\end{table*}
\endgroup
\subsection{Estimation of Time Delay for SMBHs}\label{timedel}
The time delay between different relativistic images from a time-varying light source can be measured. The total time taken by a photon travelling from a source to the observer is \cite{Bozza:2003cp,Virbhadra:2007kw}
\begin{equation}
   T=\tilde{T}\left(x_{0}\right)-\int_{D_{\mathrm{\rm OL}}}^{\rm \infty}\left|\frac{\mathrm{d} t}{\mathrm{~d} x}\right| \mathrm{d} x-\int_{D_{\mathrm{\rm LS}}}^{\rm \infty}\left|\frac{\mathrm{d} t}{\mathrm{~d} x}\right| \mathrm{d} x.\label{delay} 
\end{equation}
When both the source and the observer are located far from the black hole, the second and third terms on the right-hand side effectively cancel each other out. As a result, the total time can be expressed as: \cite{Virbhadra:2007kw}
\begin{equation}    \tilde{T}\left(x_{0}\right)=\int_{x_{0}}^{\rm \infty} \frac{2 \sqrt{B(x) C(x) A(x_{0})}}{A(x) \sqrt{\frac{C(x)}{C(x_{0})} \frac{A(x_{0})}{A(x)}-1}} \mathrm{~d} x.\label{t}
\end{equation}
The integral $\tilde T(x_0)$ represents the time the photon takes to wind around the black hole. In the strong deflection limit, the total time can be approximately  defined as a function of $u$ as \cite{Bozza:2003cp}
\begin{equation}
  \tilde{T}(u)=-\bar{a} u_{\rm ps} \log \left(\frac{u}{u_{\rm ps}}-1\right)+\bar{b} u_{\rm ps}+\mathcal{O}\left(u-u_{\rm ps}\right).\label{timedelay}
\end{equation}
If we can independently measure the time taken for the first and second image formation, the delay of the two signals $\Delta T_{2,1}$ will give an estimate of black hole parameters \cite{Bozza:2003cp}
\begin{equation}
    \Delta T^s_{2,1}=2\pi u_{\rm ps}=2\pi D_{\rm OL} \theta_\infty.\label{time}
\end{equation}
While this time delay expression (\ref{timedelay}) used the strong lensing 
approximation, it is numerically calculated by Virbhadra \cite{Virbhadra:2008ws}. In Table \ref{Table2} we compare time delay $\Delta T^s_{2,1}$ for the supermassive black holes at the centre of different galaxies representing Schwarzschild, string-inspired Euler-Heisenberg and GMGHS black holes and also calculate their difference $\delta T^s_{12}=\Delta T^s_{2,1}(Sch)-\Delta T^s_{2,1}(EH)$.

\begin{table*}
	\caption{Estimation of time delay between the first and second relativistic image for supermassive black holes at the centre of nearby galaxies considering string-inspired Euler-Heisenberg black hole ($q=0.1$ and $\eta=-1$) and GMGHS black holes ($q=0.1$ and $\eta=0$) in comparison with Schwarzschild($q\to 0$ and $\eta\to 0$). Note that time delays are calculated in minutes and $\delta T^1_{12}=\Delta T^s_{2,1}(Sch)-\Delta T^s_{2,1}(EH)$ and $\delta T^2_{12}=\Delta T^s_{2,1}(Sch)-\Delta T^s_{2,1}(GMGHS)$.}\label{Table2}
\resizebox{1\textwidth}{!}{
		\begin{tabular}{p{2cm} p{1.7cm} p{1cm} p{2.2cm} p{1.5cm} p{1.5cm} p{1.5cm} p{1.2cm} p{1.2cm}}
Galaxy & $M( M_{\odot})$ & $D_{\rm OL}$ & $M/D_{\rm OL}$  &  $\Delta T^s_{12}$ & $\Delta T^s_{12}$&$\Delta T^s_{12}$&$\delta T^1_{12}$&$\delta T^2_{12}$\\
 &  & \text{(Mpc)} &  & \text{(Sch)}& \text{(EH)}&\text{(GMGHS)} \\		
				\hline
Milky Way& $  4.3\times 10^6	 $ & $0.008 $ &       $2.471\times 10^{-11}$ & $11.4968 $ & $11.477$ & 11.4775 &$0.0198$  &0.0193   \\
M87&$ 6.5\times 10^{9} $&$ 16.8 $
&$1.758\times 10^{-11}$ & 17378.7 & 17349.8 &17349.84& 28.9&28.86 \\
NGC 4472 &$ 2.54\times 10^{9} $&$ 16.72 $
&$7.246\times 10^{-12}$& $6791.06$ &6779.78&6779.784 & 11.28&11.276 \\
			 
NGC 1332 &$ 1.47\times 10^{9} $&$22.66  $
&$3.094\times 10^{-12}$& $3930.26$ & 3923.73&3923.733 & 6.53&6.527 \\
			
NGC 4374 &$ 9.25\times 10^{8} $&$ 18.51 $
&$2.383\times 10^{-12}$& $2473.12$ & 2469.01&2469.015 &4.11&4.105\\
			 
NGC 1399&$ 8.81\times 10^{8} $&$ 20.85 $
&$2.015\times 10^{-12}$& $2355.48$ & 2351.56&2351.57 &3.92&3.91  \\
			 
NGC 3379 &$ 4.16\times 10^{8} $&$10.70$
&$1.854\times 10^{-12}$& $1112.24$ & 1110.38 &1110.39& 1.86&1.85 \\
			
NGC 4486B &$ 6\times 10^{8} $&$ 16.26 $
&$1.760\times 10^{-12}$ & $1604.19$ & 1601.52 &1601.523& 2.67&2.667 \\
			 
NGC 1374 &$ 5.90\times 10^{8} $&$ 19.57 $ &$1.438\times 10^{-12}$& $1577.45$ &  1574.83&1574.831 & 2.62&2.619 \\
			    
NGC 4649&$ 4.72\times 10^{9} $&$ 16.46 $
&$1.367\times 10^{-12}$& $12619$ &12598.64&12598.653 & 20.36&20.347\\
			
NGC 3608 &$  4.65\times 10^{8}  $&$ 22.75  $ &$9.750\times 10^{-13}$& $1243.25$ & 1241.18 &1241.185& 2.07&2.065 \\
			
NGC 3377 &$ 1.78\times 10^{8} $&$ 10.99$
&$7.726\times 10^{-13}$ & $475.909$ &  475.11&475.12 & 0.799&0.789 \\
			
NGC 4697 &$  2.02\times 10^{8}  $&$ 12.54  $ &$7.684\times 10^{-13}$& $540.077$ &  539.17&539.18 & 0.907&0.897 \\
			 
NGC 5128 &$  5.69\times 10^{7}  $& $3.62   $ &$7.498\times 10^{-13}$& $152.131$ &  151.87&151.88 & 0.261&0.251 \\
			 
NGC 1316&$  1.69\times 10^{8}  $&$20.95   $ &$3.848\times 10^{-13}$& $451.816 $ &451.09&451.095 & 0.726&0.721 \\
			 
NGC 3607 &$ 1.37\times 10^{8} $&$ 22.65  $ &$2.885\times 10^{-13}$& $366.265 $ & 365.68 &365.681& 0.585&0.584 \\
			
NGC 4473 &$  0.90\times 10^{8}  $&$ 15.25  $ &$2.815\times 10^{-13}$& $240.628$ & 240.22&240.23& 0.408&0.398\\
			
NGC 4459 &$ 6.96\times 10^{7} $&$ 16.01  $ &$2.073\times 10^{-13}$ & $186.086 $ &  185.77 &185.776& 0.316&0.31 \\

M32 &$ 2.45\times 10^6$ &$ 0.8057 $
&$1.450\times 10^{-13}$ & $6.5504 $ & 6.53 &6.539& 0.0204&0.0114 \\
Cygnus A &$2.66\times 10^9 $&$242.7$&$1.417\times10^{-15}$&7111.95&7100.08&7100.09&10.85&10.95
		\end{tabular}
}		
\end{table*}

\section{\label{constraint}EHT Observation Constraints }
\begin{figure}[!th]
    \centering
    \includegraphics[scale=.75]{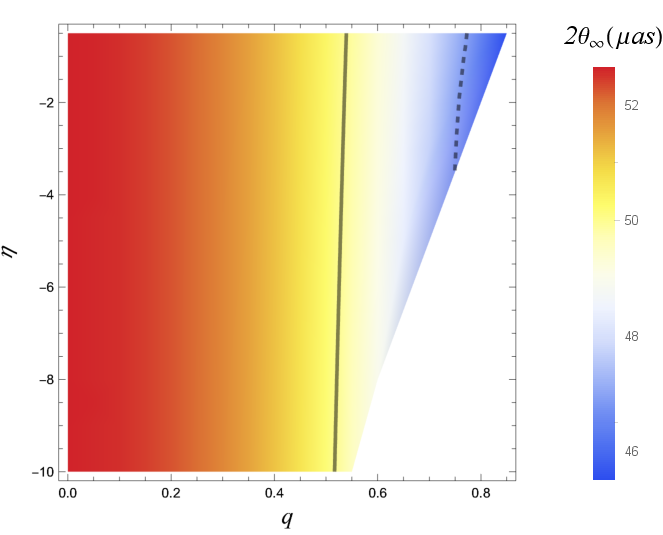}
\caption{Shadow angular diameter $\theta_{\text{sh}}(=2\theta_{\rm \infty})$ as a function of parameters $q$ and $\eta$, when Sgr A* is modelled as string-inspired Euler-Heisenberg BH. The black solid and black dashed lines correspond to $\theta_{\text{sh}}=50$ and $\theta_{\text{sh}}=46.9$, respectively. The region between these lines satisfies the Sgr A* shadow 1-$\sigma$ bound.}
\label{SgrAparameter}
\end{figure}

We consider supermassive black holes M87* and Sgr A* as gravitational lenses described by the string-inspired Euler-Heisenberg metric. We then evaluate and compare the strong-field lensing observables predicted by this metric with those corresponding to the GMGHS and Schwarzschild black holes.
The EHT collaboration observed the horizon-scaled emission from the M87* and Sgr A*, measuring the diameter of the bright emission ring $\theta_{sh}=42\pm 3~\mu$as and constraining its fractional width to be $<0.05$ and confirming the central brightness depression as the black hole's shadow signature \citep{EventHorizonTelescope:2019dse,EventHorizonTelescope:2019pgp,EventHorizonTelescope:2019ggy}. Subsequent observation in 2022 EHT revealed a thick and bright emission ring with a diameter of $51.8\pm2.3~\mu as$ \cite{EventHorizonTelescope:2022wkp}.  Using data from various imaging techniques, including EHT Imaging, SMILI, and DIFMAP—the envelope of 1-$\sigma$ interval, for the angular diameter of the shadow is $\theta_{sh} = 48.7 \pm 7~\mu$as \cite{EventHorizonTelescope:2022xqj} and quantifying the Schwarzschild shadow deviation as $\delta = -0.08^{+0.09}_{-0.09}$ (VLTI),$-0.04^{+0.09}_{-0.10}$ (Keck) \citep{EventHorizonTelescope:2022xqj}. Sgr A* presents a unique opportunity because it probes curvature orders higher than M87* and utilizes independent prior estimates for the mass-to-distance ratio. The use of three independent imaging algorithms-EHT Imaging, SMILI, and DIFMAP, ensures robustness, with the most likely values of angular shadow diameter of the Sgr A* shadow constrained to be within 46.9$\mu$as $\le$ $\theta_{\text{sh}}$ $\le$ 50 $\mu$as across different methods \citep{EventHorizonTelescope:2022xqj}. These findings not only enable exploration of strong field gravity but also offer insight into theoretical frameworks and put constraints on additional parameters within gravity theories \cite{Afrin:2021wlj,
Zakharov:2021gbg,
EventHorizonTelescope:2021dqv,
Ghosh:2020spb,Kumar:2018ple,
KumarWalia:2022aop,
Ghosh:2022kit,Kumar:2023jgh,Islam:2022wck}. Using the apparent radius of the photon sphere ($\theta_\infty$) as the angular size of the black hole shadow, we place constraints on the deviation parameters $q$ and $\eta$ by modeling M87* and Sgr A* as string-inspired Euler-Heisenberg black holes within the 1-$\sigma$ level. Then, the diameter of angular shadow ($\theta_{sh}$) is defined as $\theta_{\text{sh}}(=2\theta_{\rm \infty})$ . This supports efforts to comprehend the effects of magnetic-charged black holes in cosmic phenomena.  
\paragraph{Constraints from Sgr A*:}  We calculated shadow diameter as a function of $q$ and $\eta$ and compared it with the observed average bounds for Sgr A* shadow size $\theta_{\text{sh}} \in (46.9, 50)~\mu$as which is \cite{EventHorizonTelescope:2022xqj} in Fig.~\ref{SgrAparameter}. Here, colors represent the shadow diameter, and black solid and black dashed lines correspond to $\theta_{\text{sh}}=50\mu$as and $\theta_{\text{sh}}=46.9\mu$as respectively. While shadow size decreases with increasing $q$, it is almost independent of $\eta$.  The parameters are constrained as follows: $0.54109 \leq q \leq 0.7796$, without constraint on $\eta$. Within this defined parameter range, the string-inspired Euler-Heisenberg black hole aligns with the observations of the Sgr A* black hole shadow from the EHT. Interestingly, EHT shadow size measurement alone cannot constrain the dilaton coupling parameter $\eta$.

\begin{figure}[!th]
    \centering
    \includegraphics[scale=.75]{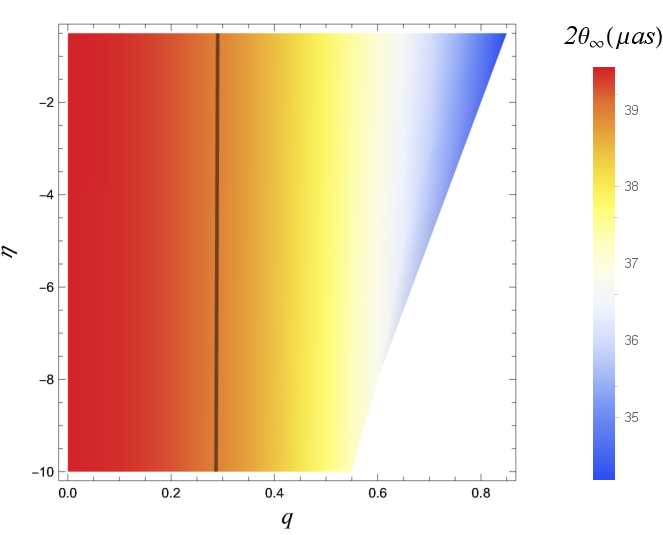}
    \caption{Shadow angular diameter $\theta_{\text{sh}}(=2\theta_{\rm \infty})$ as a function of parameters $q$ and $\eta$, when M87* is modelled as string-inspired Euler-Heisenberg BH. The black line corresponds to $\theta_{\text{sh}}=39$. The region within this line satisfies the M87* shadow 1-$\sigma$ bound.}\label{M87parameter}
\end{figure}

\paragraph{Constraints from  M87*:} In Fig.~\ref{M87parameter}, the angular diameter $\theta_{\rm sh}$ is shown as a function of $q$ and $\eta$ for the M87* black hole. Accounting for the offset of $>10\%$ between the emission ring and the angular shadow diameter, we take the mean angular diameter $\theta_sh=39~\mu as$. The black solid line represents $\theta_{\rm sh}=39\mu$as for string-inspired Euler-Heisenberg black holes as M87*. Our analysis constrains the parameters $q$ and $\eta$ such that $0< q \le 0.29107$, but there is no constraint on the value of the parameter $\eta$. Based on Fig.~\ref{M87parameter}, string-inspired Euler-Heisenberg black holes can be considered as potential candidates for astrophysical black holes.

\section{\label{conclusion}Conclusion}
We explored the strong gravitational lensing in the spacetime of spherically symmetric black holes arising in the low-energy limit of string theory, coupled with a NED field -- referred to as the string-inspired Euler-Heisenberg black hole. These black holes differ phenomenologically from the well-known GMGHS black holes due to the presence of NED fields. Specifically, we investigated the effects of string corrections ($\eta$) in the low-energy limit and the non-Maxwell NED field charge ($q$) on various strong lensing characteristics. Our findings suggest that the influence of the string coupling parameter on relativistic image positions, deflection angles, photon orbit radius, and shadow size is negligible, making it difficult to distinguish from a GMGHS black hole of the same mass and charge. Although the shadows of string-inspired Euler-Heisenberg black holes are smaller than those of GMGHS black holes, the difference in shadow size is on the order of milli-arcseconds, rendering it challenging to differentiate between the two black holes observationally.

Significant deviations from Schwarzschild lensing observables occur as the NED charge increases, compared to the string coupling constant. The deflection angle $\alpha_D$ for a fixed impact parameter $u$ is smaller in string-inspired Euler-Heisenberg black holes than in Schwarzschild black holes, further decreasing with higher NED charge $q$. Although the image positions, magnifications, and shadow size decrease with increasing charge, the separation between the outermost image and the shadow boundary slightly increases as $q$ grows. We observe that $\theta_\infty$ for Sgr A* ranges between 11.52 - 26.33 $\mu$as, with the deviation as high as 2.452 $\mu$as from Schwarzschild, while for M87* it ranges between 9.12 - 19.78  $\mu$as for M87* and the deviations is 1.89 $\mu$as. Moreover, the separation between the outermost image and the shadow boundary $s$ ranges from 32.95-39.45 $nas$ for Sgr A* while 24.75-40.2 $nas$ for M87*, respectively.  

The shadow size is nearly independent of the parameter $\eta$, meaning the EHT observations of Sgr A* and M87* provide no significant constraints on $\eta$. However, for the $\eta\leq 0$, the NED charge $q$ is constrained within the ranges $0.54109 \leq q \leq 0.7796$ for Sgr A* and $0 \leq q \leq 0.29107$ for M87*.

We found that the time delay between successive relativistic images in the string-inspired Euler-Heisenberg black hole spacetime is largely consistent with predictions from GR. For example, the difference in time delay between the Euler-Heisenberg and Schwarzschild black holes in the M87* model is approximately 28.9 minutes. In contrast, for the Sgr A* model, it is around 0.0198 minutes. Except for Sgr A*, these time-delay differences are substantial enough to be detectable with current astronomical observations, provided sufficient angular resolution between the relativistic images is achieved. Although some lensing observables deviate from GR, the overall consistency with GR predictions suggests that string-inspired Euler-Heisenberg black holes could serve as a viable model for astrophysical black hole phenomena.

\begin{appendix}
\section{Null Geodesics in String-Inspired Euler-Heisenberg Spacetime}\label{Apd-1}
Equation~(\ref{eq3}) reveals that a non-trivial coupling between the dilaton field and the NED fields modifies the photon dispersion relation and consequently their geodesic equations compared to the case with a minimally coupled Maxwell field. Following the Hadamard's approach~\cite{Hadamard} to light propagation, we deduce from Eq.~(\ref{eq3}) that photons follow null geodesics of an effective metric, $g_{a b }^{\text{(eff)}}$ that depend on the NED and dilaton fields, rather than that of the background metric $g_{a b}$. Considering $k^a$ as a null vector representing photons four-velocity, this relationship is expressed as:
\begin{eqnarray}\label{A3}
&&\Bigg(\left(2\beta f(\phi) \mathcal{F}^2-e^{-2\phi}\right)\mathcal{F}^2 g_{a b}-4\beta f(\phi)\mathcal{F}^{\mu\nu}\mathcal{F}_{\delta b}\mathcal{F}_{a\nu}\mathcal{F}^{\delta}_{\mu}\nonumber\\
&&-4\alpha\Big( \mathcal{F}_{\delta\lambda}\mathcal{F}^{\nu\lambda}\mathcal{F}_{a\nu} \mathcal{F}_{b}^{\delta}- \mathcal{F}^{\nu}_{b}\mathcal{F}_{a\nu}\mathcal{F}^2 \nonumber\\
&&- \mathcal{F}^{\mu \delta}\mathcal{F}_{\delta b}\mathcal{F}_{a \nu} \mathcal{F}_{\mu}^{\nu}\Big)
 \Bigg)k^ak^b=0.\\
&& \Rightarrow g_{a b }^{\text{(eff)}}k^ak^b=0.
\end{eqnarray}
For $\alpha = \beta = 0$, the metric~(\ref{line-elem}) reduces to the GMGHS black hole solution. In this case, photons continue to follow null geodesics of the background metric, satisfying $g_{ab} k^a k^b = 0$ and $k^a\nabla_ak^b=0$. This is evident from the effective metric in Eq.~(\ref{A3}), which reduces to the background metric $g_{ab}$ for $\alpha,\beta=0$.

From Eq.~(\ref{line-elem}), it is clear that the dilaton field's contribution to the black hole metric is of the order $\mathcal{O}(r^{-6})$ near the horizon. This indicates that the induced effects of these fields on photon geodesics are also sufficiently small, allowing us to neglect the modified dispersion relation in Eq.~(\ref{A3}) under typical astrophysical conditions. Thus, the modified null geodesic equations can be treated as small perturbations to the null geodesics of the background metric. Consequently, for this string-inspired Euler-Heisenberg black hole, photons can be reasonably approximated to follow the null geodesics of the background metric, significantly simplifying the analysis while maintaining physical accuracy within the small correction limit, consistent also with the treatment in Refs.~\cite{Jiang:2024njc, Xu:2024gjs}.

However, these corrections become substantial in regions of extreme curvature near small-$r$ or intense NED fields, such as near naked singularities, where photons originating close to $r=0$ may reach distant observers. In such cases, the modified geodesic equations governed from the effective metric (\ref{A3}) for photons must be fully incorporated. On the other hand, for supermassive black holes, where the curvature is negligible and both the NED field and the dilaton field are weak outside the horizon, deviations from null geodesic motion are expected to be minimal. This justifies the approximation for astrophysical scenarios while highlighting the necessity of accounting for modified equations in exotic regimes. 
\section{Curvature Scalar Analysis}\label{append2}
In this section, we analyze the curvature scalars of string-inspired Euler-Heisenberg black holes given by metric (\ref{5}). In terms of ADM mass $M$ of the black hole and radial coordinate $r$, the black hole's metric (\ref{5}) can be written as:
\begin{equation}
{\rm d}\tilde{s}^2=-A(r){\rm d}t^2+B(r){\rm d}r^2+C(r)({\rm d}\vartheta^2+\sin^2\vartheta d\varphi^2),
\label{ametric}
\end{equation}
where
\begin{equation}
\begin{aligned}
A(r)&=1-\frac{4M^2}{Q^2+\sqrt{Q^4 + 4r^2M^2}}-\frac{2(\alpha-\beta)  Q^4 }{r^6},\\
B(r)&=\frac{4r^2M^2}{(Q^4 + 4r^2M^2)A(r)},~~~~ C(r)=r^2,
\end{aligned}
\label{afunction}
\end{equation}
 Now the scalar curvature \( R \) for (\ref{ametric}) is given by:
\begin{eqnarray}
&R = 
\frac{Q^4 \left(Q^2 r^4 + r^6 - r^4 \sqrt{Q^4 + 4 M^2 r^2} + 22 Q^4 (\alpha - \beta) + 80 M^2 r^2 (\alpha - \beta)\right)}{2 M^2 r^{10}}\nonumber\\
\label{scalar}   
\end{eqnarray}

, and The Kretschmann invariant \( K \) is given by:
\begin{widetext}
\begin{eqnarray}
\nonumber
&K = \frac{1}{4 M^4 r^{20}} \Biggl(
192 M^6 r^{14} - 192 M^4 Q^2 r^{12} \sqrt{Q^4 + 4 M^2 r^2} 
+ 16 M^2 Q^4 r^8 \left(-r^4 \sqrt{Q^4 + 4 M^2 r^2} + M^2 \left(33 r^4 \right.\right. \\\nonumber
&\left.\left. + 112 \sqrt{Q^4 + 4 M^2 r^2} (\alpha - \beta)\right)\right) 
- 8 M^2 Q^6 r^8 \left(-3 r^4 + 27 r^2 \sqrt{Q^4 + 4 M^2 r^2} + 608 M^2 (\alpha - \beta)\right) \\\nonumber
&+ 2508 Q^{16} (\alpha - \beta)^2 
- 476 Q^{14} r^4 (\alpha - \beta) 
+ 2 Q^{12} r^2 \left(27 r^6 + 238 r^2 \sqrt{Q^4 + 4 M^2 r^2} (\alpha - \beta) \right. \\
&\left. + 8752 M^2 (\alpha - \beta)^2 - 18 r^4 (\alpha - \beta)\right) 
+ 2 Q^{10} r^6 \left(-27 r^2 \sqrt{Q^4 + 4 M^2 r^2} + 5 \left(r^4 - 304 M^2 (\alpha - \beta)\right)\right) \\\nonumber
&+ Q^8 r^4 \biggl(3 r^8 - 10 r^6 \sqrt{Q^4 + 4 M^2 r^2} + 30592 M^4 (\alpha - \beta)^2 \\\nonumber
&\quad + 4 M^2 r^2 \left(81 r^4 + 540 \sqrt{Q^4 + 4 M^2 r^2} (\alpha - \beta) - 28 r^2 (\alpha - \beta)\right)\biggr)
\Biggr).\label{kre}  
\end{eqnarray}
\end{widetext}
From the curvature scalar analysis, we see that the curvature scalar $R$ and Kretschmann invariant \( K \), diverge as $r\to0$. Hence, it possesses a single spacetime singularity residing at $r=0$

\end{appendix}

 \section{Acknowledgments} 
A.V. and S.G.G. would like to thank SERB-DST for project No. CRG/2021/005771. RKW's research is supported by the Fulbright-Nehru Postdoctoral Research Fellowship (award 2847/FNPDR/2022)  from the United States-India Educational Foundation.
\bibliography{reference1.bib}
\end{document}